# Megaprojects and Risk

*Megaprojects and Risk* provides the first detailed examination of the phenomenon of megaprojects. It is a fascinating account of how the promoters of multibillion-dollar megaprojects systematically and self-servingly misinform parliaments, the public and the media in order to get projects approved and built. It shows, in unusual depth, how the formula for approval is an unhealthy cocktail of underestimated costs, overestimated revenues, undervalued environmental impacts and overvalued economic development effects. This results in projects that are extremely risky, but where the risk is concealed from MPs, taxpayers and investors. The authors not only explore the problems but also suggest practical solutions drawing on theory and hard, scientific evidence from the several hundred projects in twenty nations that illustrate the book. Accessibly written, it will be essential reading in its field for students, scholars, planners, economists, auditors, politicians, journalists and interested citizens.

BENT FLYVBJERG is Professor in the Department of Development and Planning at Aalborg University, Denmark, and author of the highly successful *Making Social Science Matter* (Cambridge, 2001) and *Rationality and Power* (1998).

NILS BRUZELIUS is Associate Professor at Stockholm University and an independent consultant on transport and planning.

WERNER ROTHENGATTER is Head of the Institute of Economic Policy Research and of the Unit on Transport and Communication at the University of Karlsruhe, Germany.

# Megaprojects and Risk

*An Anatomy of Ambition*

Bent Flyvbjerg
Nils Bruzelius
Werner Rothengatter

CAMBRIDGE
UNIVERSITY PRESS

# Contents

# Figures

# Tables

# Acknowledgements

We wish to thank the people and organisations who helped make this book possible. Special thanks must be given to Patrick Ponsolle and John Noulton, of Eurotunnel, Mogens Bundgaard-Nielsen, of Sund & Bælt Holding, and Ole Zacchi, of the Danish Ministry of Transport. Not only did they and their staff help us with data for the book's case studies, they also gave critical comments on an earlier version of the book's manuscript.

We also wish to thank Martin Wachs, of the University of California at Berkeley, and Don Pickrell, of the Volpe National Transportation Systems Center at Cambridge, Massachusetts, for their comments on our analysis of cost overrun. Per Homann Jespersen, of Roskilde University, provided helpful input to our considerations regarding environmental impacts and risks. Roger Vickerman, of the University of Kent at Canterbury, gave valuable comments on the chapter about regional and economic growth effects. Thanks are due as well to the following colleagues for their kind help at various stages in the research and writing process: Jim Bohman, Irene Christiansen, John Dryzek, Raphael Fischler, Ralph Gakenheimer, Maarten Hajer, Mette Skamris Holm, Andy Jamison, Bill Keith, Finn Kjærsdam, Mary Rose Liverani, Kim Lynge Nielsen, Tim Richardson, Yvonne Rydin, Ed Soja, Michael Storper, Andy Thornley, Jim Throgmorton and Alan Wolfe. Two anonymous Cambridge University Press reviewers provided highly useful comments for preparing the final version of the typescript.

The transport sector and its institutions are hardly in the vanguard regarding freedom of information. In some cases we were unable, using the formal channels for information gathering, to get the data and in-depth information we needed to write the book the way we wanted to write it. We are grateful to those bold individuals who, when the formal channels dried up, found informal ways of furnishing us with the information we lacked. We mention no names for obvious reasons.

Lilli Glad expertly transformed our drafts into readable manuscripts. Anni Busk Nielsen provided precious help in acquiring the literature on which the study is based. The research and the book were made possible by generous grants from the Danish Transport Council and Aalborg University. Finally, we wish to thank our editor at Cambridge University Press, Sarah Caro, who provided valuable help in seeing the book through the printing process. Bent Flyvbjerg was teamleader for the research on which the book is based and is principal author of the book. We apologise to anyone we have forgotten to mention here. Responsibility for errors or omissions in this book remains ours alone.

# 1 The megaprojects paradox

### A new animal

Wherever we go in the world, we are confronted with a new political and physical animal: the multibillion-dollar mega infrastructure project. In Europe we have the Channel tunnel, the Øresund bridge between Denmark and Sweden, the Vasco da Gama bridge in Portugal, the German MAGLEV train between Berlin and Hamburg, the creation of an interconnected high-speed rail network for all of Europe, cross-national motorway systems, the Alp tunnels, the fixed link across the Baltic Sea between Germany and Denmark, plans for airports to become gateways to Europe, enormous investments in new freight container harbours, DM 200 billion worth of transport infrastructure projects related to German unification alone, links across the straits of Gibraltar and Messina, the world's longest road tunnel in Norway, not to speak of new and extended telecommunications networks, systems of cross-border pipelines for transport of oil and gas, and cross-national electrical power networks to meet the growing demand in an emerging European energy market. It seems as if every country, and pair of neighbouring countries, is in the business of promoting this new animal, the megaproject, on the European policy-making scene. And the European Union, with its grand scheme for creating so-called 'Trans-European Networks', is an ardent supporter and even initiator of such projects, just as it is the driving force in creating the regulatory, and de-regulatory, regimes that are meant to make the projects viable.[1]

The situation is similar in industrialised and industrialising countries in other parts of the world, from Asia to the Americas. There is, for example, Hong Kong's Chek Lap Kok airport, China's Quinling tunnel, the Akashi Kaikyo bridge in Japan, Sydney's harbour tunnel, Malaysia's North–South Expressway, Thailand's Second Stage Expressway, and proposals for an integrated Eurasian transport network. In the Americas there is Boston's 'Big Dig', freeways and railways in California, Denver's new international airport, Canada's Confederation bridge, the São Paulo–Buenos Aires Superhighway, the Bi-Oceanic highway right across South America from the Atlantic to the Pacific, and the Venezuela–Brazil highway. Even a proposed US$50 billion project to link the USA and Russia across the Bering Strait – the 'biggest project in history', according to its promoters – is not missing in the megaproject scheme of things.[2] Outside the field of transport infrastructure there is the Three Gorges dam in China, Russia's natural gas pipelines, the Pergau dam in Malaysia, flood control in Bangladesh, the Bolivia–Brazil gas pipeline, the Venezuela–Brazil power line and, again and everywhere, the ultimate megaproject, the Internet with associated infrastructure and telecommunications projects.

> Megaprojects form part of a remarkably coherent story, the 'Great War of Independence from Space'.

### Zero-friction society

Megaprojects form part of a remarkably coherent story. Sociologist Zygmunt Bauman perceptively calls it the 'Great War of Independence from Space', and he sees the resulting new mobility as the most powerful and most coveted stratifying factor in contemporary society.[3] Paul Virilio speaks of the 'end of geography' while others talk of the 'death of distance'.[4] Bill Gates, founder and chair of Microsoft Corporation, has dubbed the phenomenon 'frictionless capitalism' and sees it as a novel stage in capitalist evolution.[5] When Microsoft and Gates single out a concept or a product, one is well advised to pay attention. 'Frictionless society' may sound like an advertiser's slogan in the context of its usage. It is not. The term signifies a qualitatively different stage of social and economic development.

In this development 'infrastructure' has become a catchword on a par with 'technology'. Infrastructure has rapidly moved from being a simple precondition for production and consumption to being at the very core of these activities, with just-in-time delivery and instant Internet access being two spectacular examples of this. Infrastructure is the great space shrinker, and power, wealth and status increasingly belong to those who know how to shrink space, or know how to benefit from space being shrunk.[6]

Today infrastructure plays a key role in nothing less than the creation of what many see as a new world order where people, goods, energy,

information and money move about with unprecedented ease. Here the politics of distance is the elimination of distance. The name of utopia is Zero-Friction Society. And even if we can never achieve utopian frictionlessness, we may get close, as is currently happening with the spread of the Internet. Modern humans clearly have a preference for independence from space and are consistently undercutting the friction of distance by building more and improved infrastructure for transport, including telecommunications and energy.

Megaprojects are central to the new politics of distance because infrastructure is increasingly being built as megaprojects. Thus the past decade has seen a sharp increase in the magnitude and frequency of major infrastructure projects, supported by a mixture of national and supranational government, private capital and development banks.

> Many projects have strikingly poor performance records in terms of economy, environment and public support.

### Performance paradox

There is a paradox here, however. At the same time as many more and much larger infrastructure projects are being proposed and built around the world, it is becoming clear that many such projects have strikingly poor performance records in terms of economy, environment and public support.[7] Cost overruns and lower-than-predicted revenues frequently place project viability at risk and redefine projects that were initially promoted as effective vehicles to economic growth as possible obstacles to such growth. The Channel tunnel, opened in 1994 at a construction cost of £4.7 billion, is a case in point, with several near-bankruptcies caused by construction cost overruns of 80 per cent, financing costs that are 140 per cent higher than those forecast and revenues less than half of those projected (see Chapters 2–4). The cost overrun for Denver's US$5 billion new international airport, opened in 1995, was close to 200 per cent and passenger traffic in the opening year was only half of that projected. Operating problems with Hong Kong's new US$20 billion Chek Lap Kok airport, which opened in 1998, initially caused havoc not only to costs and revenues at the airport; the problems spread to the Hong Kong economy as such with negative effects on growth in gross domestic product.[8] After nine months of operations, *The Economist* dubbed the airport a 'fiasco', said to have cost the Hong Kong economy US$600 million.[9] The fiasco may have been only a start-up problem, albeit an expensive one, but it

is the type of expense that is rarely taken into account when planning megaprojects.

Some may argue that in the long term cost overruns do not really matter and that most monumental projects that excite the world's imagination had large overruns. This line of argument is too facile, however. The physical and economic scale of today's megaprojects is such that whole nations may be affected in both the medium and long term by the success or failure of just a single project. As observed by Edward Merrow in a RAND study of megaprojects:

> Such enormous sums of money ride on the success of megaprojects that company balance sheets and even government balance-of-payments accounts can be affected for years by the outcomes... The success of these projects is so important to their sponsors that firms and even governments can collapse when they fail.[10]

Even for a large country such as China, analysts warn that the economic ramifications of an individual megaproject such as the Three Gorges dam 'could likely hinder the economic viability of the country as a whole'.[11] Stated in more general terms, the Oxford-based Major Projects Association, an organisation of contractors, consultants, banks and others interested in megaproject development, in a recent publication speaks of the 'calamitous history of previous cost overruns of very large projects in the public sector'. In another study sponsored by the Association the conclusion is, 'too many projects proceed that should not have done'.[12] We would add to this that regarding cost overruns there is no indication that the calamity identified by the Major Projects Association is limited to the public sector. Private sector cost overruns are also common.

For environmental and social effects of projects, one similarly finds that such effects often have not been taken into account during project development, or they have been severely miscalculated.[13] In Scandinavia, promoters of the Øresund and Great Belt links at first tried to ignore or downplay environmental issues, but were eventually forced by environmental groups and public protest to accept such issues on the decision-making agenda (see Chapter 5). In Germany, high-speed rail projects have been criticised for not considering environmental disruption. Dams are routinely criticised for the same thing. However, environmental problems that are not taken into account during project preparation tend to surface during construction and operations; and such problems often destabilise habitats, communities and megaprojects themselves, if not dealt with carefully. Moreover, positive regional development effects, typically much touted by project promoters to gain political acceptance for their projects, repeatedly turn out to be non-measurable, insignificant or even negative (see Chapter 6).

In consequence, the cost–benefit analyses, financial analyses and environmental and social impact statements that are routinely carried out as part of megaproject preparation are called into question, criticised and denounced more often and more dramatically than analyses in any other professional field we know. Megaproject development today is not a field of what has been called 'honest numbers'.[14] It is a field where you will see one group of professionals calling the work of another not only 'biased' and 'seriously flawed' but a 'grave embarrassment' to the profession.[15] And that is when things have not yet turned unfriendly. In more antagonistic situations the words used in the mud-slinging accompanying many megaprojects are 'deception', 'manipulation' and even 'lies' and 'prostitution'.[16] Whether we like it or not, megaproject development is currently a field where little can be trusted, not even – some would say especially not – numbers produced by analysts.

Finally, project promoters often avoid and violate established practices of good governance, transparency and participation in political and administrative decision making, either out of ignorance or because they see such practices as counterproductive to getting projects started. Civil society does not have the same say in this arena of public life as it does in others; citizens are typically kept at a substantial distance from megaproject decision making. In some countries this state of affairs may be slowly changing, but so far megaprojects often come draped in a politics of mistrust. People fear that the political inequality in access to decision-making processes will lead to an unequal distribution of risks, burdens and benefits from projects.[17] The general public is often sceptical or negative towards projects; citizens and interest groups orchestrate hostile protests; and occasionally secret underground groups even encourage or carry out downright sabotage on projects, though this is not much talked about in public for fear of inciting others to similar guerrilla activities.[18] Scandinavians, who like other people around the world have experienced the construction of one megaproject after another during the past decade, have coined a term to describe the lack in megaproject decision making of accustomed transparency and involvement of civil society: 'democracy deficit'. The fact that a special term has come into popular usage to describe what is going on in megaproject decision making is indicative of the extent to which large groups in the population see the current state of affairs as unsatisfactory.

> Civil society does not have the same say in this arena of public life as it does in others. Megaprojects often come draped in a politics of mistrust.

### Risk, democracy and power

The megaprojects paradox consists in the irony that more and more megaprojects are built despite the poor performance record of many projects. In this book we link the idea of megaprojects with the idea of risk and we identify the main causes of the megaprojects paradox to be inadequate deliberation about risk and lack of accountability in the project decision-making process. We then proceed to propose ways out of the paradox. We will show that in terms of risk, most appraisals of megaprojects assume, or pretend to assume, that infrastructure policies and projects exist in a predictable Newtonian world of cause and effect where things go according to plan. In reality, the world of megaproject preparation and implementation is a highly risky one where things happen only with a certain probability and rarely turn out as originally intended.

Sociologists such as Ulrich Beck and Anthony Giddens have argued that in modern society risk has increasingly become central to all aspects of human affairs; that we live in a 'risk society' where deliberation about social, economic, political and environmental issues is bound to fail if it does not take risk into account.[19] If this diagnosis is correct – and we will argue that for megaprojects it is – then it is untenable to continue to act as if risk does not exist or to underestimate risk in a field as costly and consequential as megaproject development.

The Beck–Giddens approach to risk society is our point of departure for understanding risk and its particular relevance to modern society. Yet this approach does not take us far enough in the direction we want to go. The problem with Beck, Giddens and related theories is that they use risk mainly as a metaphor for mature modernity. We want to proceed beyond the level of symbol and theory to use risk as an analytic frame and guide for actual decision making. We will do this by developing a set of ideas of how risk assessment and risk management may be used as vehicles for governing risk.[20] In the words of Silvio Funtowicz and Jerome Ravetz, where facts are uncertain, decision-stakes high and values in dispute, risk assessment must be at the heart of decision making.[21] A growing number of society's decision areas meet these criteria. Megaproject development is one of them.

We do not believe risk can be eliminated from risk society. We believe, however, that risk may be acknowledged much more explicitly and managed a great deal better, with more accountability, than is typically the case today. Like Ortwin Renn, Thomas Webler and others, we hold that risk assessment and management should involve citizens and stakeholders to reflect their experience and expertise, in addition to including the usual suspects, namely government experts, administrators and politicians.[22]

We here define stakeholders as key institutional actors, such as NGOs, various levels of government, industrial interests, scientific and technical expertise and the media. Some of these stakeholder groups will claim to be speaking legitimately on behalf of the public good and some, but not all of them, will be doing so. Given that such stakeholders do not always adequately represent publics, we recognise the need, on both democratic and pragmatic grounds, to properly involve publics in decision making. Such involvement should take place in carefully designed deliberative processes from the beginning and throughout large-scale projects.[23] Like Renn and Webler, we believe that one should go as far as possible with the participatory and deliberative approach in including publics and stakeholders and that the result will be decisions about risk that are better informed and more democratic.

We find, nevertheless, that deliberative approaches to risk, based as they are on communicative rationality and the goodwill of participants, can take us only some of the way towards better decisions and will frequently fail for megaprojects.[24] This is so because the interests and power relations involved in megaprojects are typically very strong, which is easy to understand given the enormous sums of money at stake, the many jobs, the environmental impacts, the national prestige, and so on. Communicative and deliberative approaches work well as ideals and evaluative yardsticks for decision making, but they are quite defenceless in the face of power.[25] And power play, instead of commitment to deliberative ideals, is often what characterises megaproject development. In addition to deliberative processes, we also focus, therefore, on how power relations and outcomes may be influenced and balanced by reforming the institutional arrangements that form the context of megaproject decision making.[26]

Based on this approach to risk, it is an essential notion of the book that good decision making is a question not only of better and more rational information and communication, but also of institutional arrangements that promote accountability, and especially accountability towards risk. We see accountability as being a question not just about periodic elections, but also about a continuing dialogue between civil society and policy makers and about institutions holding each other accountable through appropriate checks and balances.[27] Thus we replace the conventional decisionistic approach to megaproject development with a more current institutionalistic one centred on the practices and rules that comprise risk and accountability.[28] We also hold that our approach must be based on actual experience from concrete projects. The purpose is to ensure a realistic understanding of the issues at hand as well as proposals that are practically desirable and possible to implement.

### A brief overview

We build our case for a new approach to megaproject decision making in two main steps. In the first half of the book, we identify the weaknesses of the conventional approach to megaproject development. By so doing we argue that a different approach is needed. Our critique of the conventional approach is proactive; from the critique we tease out the problems, namely problems that need to be embraced by an alternative approach. In the second half of the book, we argue empirically and theoretically how the weaknesses of the conventional approach can be overcome by emphasising risk, institutional issues and accountability. Finally, an example for readers with a practical bent is included in the appendix, which shows how our approach to megaproject decision making was employed for a specific project with which we have been involved as advisers to the Danish government, namely the proposed Baltic Sea link connecting Germany and Denmark across Fehmarn Belt, one of the largest cross-national infrastructure projects in the world.

Throughout the book we illustrate major points on the basis of in-depth case studies of three recent megaprojects that form part of the so-called Trans-European Transport Network sponsored by the European Union and national governments:

(1) the Channel tunnel, also known as the 'Chunnel', between France and the UK, which opened in 1994 and is the longest underwater rail tunnel in Europe;
(2) the Great Belt link, opened in 1997–98, connecting East Denmark with continental Europe, and including the longest suspension bridge in Europe plus the second longest underwater rail tunnel; and finally
(3) the Øresund link between Sweden and Denmark, which opened in 2000, and which connects the rest of Scandinavia with continental Europe.

The three case studies are supplemented by material from a large number of other major projects, mainly from the field of transport infrastructure, but also from other fields such as information technology, power plants, water projects, oil and gas extraction projects and aerospace projects. The economics and politics of building a bridge or an airport are surely different on many points from those of space exploration, water management, or providing global access to the Internet. But despite such differences, our data show that there are also important similarities, for instance regarding cost overrun and financial risk, where we find a remarkably similar pattern across different project types. We argue that the measures of accountability that are necessary for detecting and curbing systematic cost underestimation, benefit overestimation and other risks are quite similar across projects. Thus, even though the main focus of the book is the development of mega transport infrastructure projects, the approach developed is relevant for other types of megaproject as well.

Our case studies and other data cover both public and private sector projects. We argue that for megaprojects there is no simple formula for the government–business divide. Megaprojects are so complicated that by nature they are essentially hybrid. This is the case even for projects that are considered fully private, for instance the Channel tunnel, because the sheer complexity and potential impacts of a megaproject dictate deep public-sector involvement for many issues, for instance regarding safety and environment. Thus public–private collaboration is crucial, even for private-sector projects. The question is not whether such collaboration should take place but how. In Chapters 9 and 10 we address this question and redraw the borderlines of public and private involvement in megaproject development with a view to improving governance of risk.

By linking the idea of megaprojects with the idea of risk we hope to broaden the scope of the risk literature and to attract attention to this topic. As far as we are aware, no other study does this today. In writing the book, we have aimed at an interdisciplinary audience of students and scholars in the social and decision-making sciences with an interest in risk, public policy and planning, ranging from sociology and social policy to political science and public policy to public administration, management and planning. Policy makers, administrators and planners are also an important target group for the book, as are consultants, auditors and other practitioners working with megaproject development. We maintain that governments and developers who continue to ignore the type of knowledge and proposals presented here do so at their own peril. Megaprojects are increasingly becoming highly public and intensely politicised ventures drawing substantial international attention with much potential for generating negative publicity.

The Three Gorges dam mentioned above is a case in point. So is the 650 km Myanmar–Thailand natural gas pipeline and maintenance road, built through pristine natural forests and habitats. Lonely Planet, the world's leading travel guidebook publisher, decided to print, up front in its best-selling Thailand guide, a highly visible protest against the pipeline which called the actions of both the Thai government and named transnational companies, such as American Unocal and French Total, a 'scam', 'shameful' and a '*fait accompli*'.[29] Lonely Planet encourages the reader to join the protests against the project and lists – thorough as always – three addresses and telephone and fax numbers where that is possible. This is hardly how the Tourist Authority of Thailand would have preferred to present the country to its visitors, nor is it the type of publicity that transnational corporations opt for, if they have a choice. Our point is they do: there is another way to deal with megaprojects and this book explains what this is.

Finally, though we did not write the book with lay readers primarily in mind, we hope that individuals, communities, activists, media and the general public interested in and affected by megaproject development will find useful insights in the book, for instance regarding the deceptions and power games they are likely to encounter if they get involved with megaprojects. Understanding the anatomy of megaprojects is necessary to be an effective player in project development. And, as mentioned, we see stronger involvement by civil society and stakeholder groups in megaproject decision making as a prerequisite for decisions that are better informed and more democratic.

Theorists of risk society and democracy have recently begun to contemplate the type of practical policy and planning needed for dealing with risk in real-life public deliberation and decision making. 'In risk society', one study concludes, 'public policy requires long-term planning for uncertainty, within a clear framework of principles and evidence to support devolved and flexible decision making. This, in turn, requires the involvement of informed and active citizens, enjoying a mature, adult-to-adult relationship with experts and with politicians. A high-trust democracy: the only way to face a risky future.'[30] In order for this approach to work, the trust in 'high-trust democracy' must be based on, not feet-in-the-air idealism about the merits of democracy, but hard-nosed considerations about risk and democratic accountability. Life will never be risk free, we are happy to report. But risk can be faced in ways much more intelligent than those currently seen. We offer this book as an attempt at fleshing out in practice the type of decision making and democracy called for by theorists of risk and democracy for a specific domain of increasing social, economic and political importance, namely that of megaproject development.

# Notes

1 THE MEGAPROJECTS PARADOX

1. On the role of the European Union as a promoter of megaprojects, see John F. L. Ross, *Linking Europe: Transport Policies and Politics in the European Union* (Westport, CT: Praeger Publishers, 1998). See also OECD, *Infrastructure Policies for the 1990s* (Paris: OECD, 1993); and Roger W. Vickerman, 'Transport Infrastructure and Region Building in the European Community', *Journal of Common Market Studies*, vol. 32, no. 1, March 1994, pp. 1–24.
2. *The Economist*, 19 August 1995, p. 84.
3. Zygmunt Bauman, *Globalization: The Human Consequences* (Cambridge: Polity Press, 1998); here quoted from Bauman, 'Time and Class: New Dimensions of Stratification', *Sociologisk Rapportserie*, no. 7, Department of Sociology, University of Copenhagen, 1998, pp. 2–3.
4. Paul Virilio, 'Un monde surexposé: fin de l'histoire, ou fin de la géographie?', in *Le Monde Diplomatique*, vol. 44, no. 521, August 1997, p. 17, here quoted from Bauman 'Time and Class'. According to Bauman, the idea of the 'end of geography' was first advanced by Richard O'Brien, in *Global Financial Integration: The End of Geography* (London: Chatham House/Pinter, 1992). See Frances Cairncross, *The Death of Distance: How the Communications Revolution Will Change Our Lives* (Boston, MA: Harvard Business School Press, 1997). See also Linda McDowell, ed., *Undoing Place? A Geographical Reader* (London: Arnold, 1997).
5. *Time*, 3 August 1998.
6. Although dams are not part of transport and communication infrastructure as such, we consider the building of dams to be part of the war of independence from space. Dams typically involve the production of electricity and electricity is one of the most effective ways of freeing industry from localised sources of energy and thus for making industry 'footloose', i.e. independent from space.
7. Peter W. G. Morris and George H. Hough, *The Anatomy of Major Projects: A Study of the Reality of Project Management* (New York: John Wiley & Sons, 1987); Mads Christoffersen, Bent Flyvbjerg and Jørgen Lindgaard Pedersen, 'The Lack of Technology Assessment in Relation to Big Infrastructural Decisions', in *Technology and Democracy: The Use and Impact of Technology Assessment in Europe. Proceedings from the 3rd European Congress on Technology Assessment*, vol. I, Copenhagen: n. p., 4–7 November 1992, pp. 54–75; David Collingridge, *The Management of Scale: Big Organizations, Big Decisions, Big*

18. Brian Doherty, 'Paving the Way: The Rise of Direct Action Against Road-Building and the Changing Character of British Environmentalism', *Political Studies*, vol. 47, no. 2, June 1999, pp. 275–91; Andrea D. Luery, Luis Vega and Jorge Gastelumendi de Rossi, *Sabotage in Santa Valley: The Environmental Implications of Water Mismanagement in a Large-Scale Irrigation Project in Peru* (Norwalk, CT: Technoserve, 1991); Jon Teigland, 'Predictions and Realities: Impacts on Tourism and Recreation from Hydropower and Major Road Developments', *Impact Assessment and Project Appraisal*, vol. 17, no. 1, March 1999, p. 67; 'Svensk webbsida uppmanar till sabotage' (Swedish website is encouraging sabotage) and 'Sabotage för miljoner' (sabotage for millions), *Svensk Vägtidning*, vol. 84, no. 2, 1997, p. 3 and vol. 85, no. 1, 1998, p. 7. One of the authors of the present book has similarly come across sabotage of a large-scale irrigation project in the Kilimanjaro region in Tanzania: see Bent Flyvbjerg, *Making Social Science Matter: Why Social Inquiry Fails and How It Can Succeed Again* (Cambridge: Cambridge University Press, 2001), chap. 10.
19. Ulrich Beck, *Risk Society: Towards a New Modernity* (Thousand Oaks, CA: Sage, 1992); Anthony Giddens, *The Consequences of Modernity* (Stanford, CA: Stanford University Press, 1990); Jane Franklin, ed., *The Politics of Risk Society* (Cambridge: Polity Press, 1998).
20. For an introduction to the literature on risk assessment and management, see Sheldon Krimsky and Dominic Golding, eds., *Social Theories of Risk* (Westport, CT: Praeger, 1992); Ortwin Renn, 'Three Decades of Risk Research: Accomplishments and New Challenges', *Journal of Risk Research*, vol. 1, no. 1, 1998, pp. 49–71. See also Chapter 7 and the three key journals in the field, *Journal of Risk Research*, *Risk Analysis* and *Journal of Risk and Uncertainty*.
21. Silvio O. Funtowicz and Jerome R. Ravetz, 'Three Types of Risk Assessment and the Emergence of Post-normal Science', in Krimsky and Golding, eds., *Social Theories of Risk*, pp. 251–73. See also Carlo Jaeger, Ortwin Renn, Eugene A. Rosa and Thomas Webler, *Risk, Uncertainty and Rational Action* (London: Earthscan, 2001).
22. Ortwin Renn, Thomas Webler and Peter Wiedemann, eds., *Fairness and Competence in Citizen Participation: Evaluating Models for Environmental Discourse* (Dordrecht: Kluwer, 1995); Ortwin Renn, 'A Model for an Analytic-Deliberative Process in Risk Management', *Environmental Science and Technology*, vol. 33, no. 18, September 1999, pp. 3049–55; Thomas Webler and Seth Tuler, 'Fairness and Competence in Citizen Participation: Theoretical Reflections From a Case Study', *Administration and Society*, vol. 32, no. 5, November 2000, pp. 566–95.
23. Adolf G. Gundersen, *The Environmental Promise of Democratic Deliberation* (Madison, WI: University of Wisconsin Press, 1995); Katherine E. Ryan and Lizanne Destefano, eds., *Evaluation As a Democratic Process: Promoting Inclusion, Dialogue, and Deliberation* (San Francisco: Jossey-Bass, 2000); Edward C. Weeks, 'The Practice of Deliberative Democracy: Results From Four Large-Scale Trials', *Public Administration Review*, vol. 60, no. 4, July–August

30. Anna Coote, 'Risk and Public Policy: Towards a High-Trust Democracy', in Jane Franklin, ed., *The Politics of Risk Society* (Cambridge: Polity Press, 1998), p. 131.

2 A CALAMITOUS HISTORY OF COST OVERRUN

1. Information from The Channel Tunnel Group, July 1998, kindly made available by Mette K. Skamris, Department of Development and Planning, Aalborg University. According to Eurotunnel, the 1986 Prospectus made provision for a standby loan facility of one billion pounds to provide for such contingencies as delays, additional capital expenditures, etc. The above cost figure does not include this contingency, according to Eurotunnel (correspondence with Eurotunnel, December 1999, authors' archives).
2. Sund & Bælt, *Årsberetning 1999* (Copenhagen: Sund & Bælt Holding, 2000), p. 19.
3. Danish Parliament, 'Bemærkninger til Forslag til Lov om anlæg af fast forbindelse over Øresund', Lovforslag nr. L 178 (Folketinget [Danish Parliament] 1990–91, 2. samling, proposed 2 May 1991), p. 10.
4. Sund & Bælt, *Årsberetning 1999*, p. 20.
5. Sund & Bælt, *Årsberetning 1999*, p. 20; Danish Parliament, 'Redegørelse af 6/12 93 om anlæg af fast forbindelse over Øresund', Fortryk af Folketingets forhandlinger, Folketinget, 7 December 1993 (Copenhagen: sp. 3212–3213); Danish Auditor-General, *Beretning til statsrevisorerne om udviklingen i de økonomiske overslag vedrørende Øresundsforbindelsen* (Copenhagen: Rigsrevisionen (Danish Auditor-General), November 1994), pp. 43–44. Øresundskonsortiet, *Den faste forbindelse over Øresund* (Copenhagen: Øresundskonsortiet, 1994), p. 4.
6. In a response to our description of cost development in the Great Belt and Øresund links, the management of the two links objects to our figures for cost overrun including cost increases that were caused by politically approved changes to project designs, with a view, for instance, to environmental protection. According to the management, such cost increases should not be included in the figures for cost overrun. Also, the management comments that the baseline budget used for calculating cost overrun should not be the budget at the time of decision to build but a later budget estimated after the politically approved design changes and after the project companies had been established and had taken over responsibility for the projects (correspondence with Sund & Bælt Holding, 20 December 1999, authors' archives). If we followed the recommendations of the management, the result would be cost overruns substantially lower than those mentioned in the main text. While we understand why the project management, from their point of view, would prefer not to include in the calculation of cost overrun items and time periods for which they were not responsible, we maintain that the internationally accepted standard for calculating cost overrun is to compare actual costs with costs estimated at the time of decision to build. This is the standard for good reason. First, the information available to those making the decision to build, at the time they make it, is what is relevant when we want to evaluate whether the decision was an informed one or not. Second, this standard for calculating cost overrun